\documentclass[aps,12pt,showpacs]{revtex4}
\usepackage{color}
\usepackage{amsmath}
\usepackage{amstext}
\usepackage{amssymb}
\usepackage[latin2]{inputenc}
\newcommand{\gnot}{{\rm GCNOT }}
\newcommand{\cnot}{{\rm CNOT }}
\newcommand{\dft}{{\rm DFT }}
\newcommand{\omd}{\omega_{d}}

\begin{document}

\author{Antoni W\'ojcik}
\email{antwoj@amu.edu.pl}

\author{Andrzej Grudka}
\author{Ravindra W.  Chhajlany}
\affiliation{Faculty of Physics, Adam Mickiewicz University,
  Umultowska 85, 61-614 Pozna\'n, Poland}
\title{Generation of inequivalent generalized  Bell bases}

\begin{abstract}
  The notion of equivalence of maximally entangled bases of bipartite
  {\it d}--dimensional Hilbert spaces ${\cal H}_{d} \otimes {\cal
    H}_{d}$ is introduced.  An explicit method of inequivalent bases
  construction is presented.
\end{abstract}
\maketitle

Maximally entangled states of two qubits - the so called Bell states
\cite{braun+92} form a basic element of many quantum information
protocols - e.g.  quantum teleportation \cite{ben+93} and quantum
dense coding \cite{benwies92} to mention only the most popular.
Recently some protocols have been generalized to make use of {\it
  d}--dimensional quantum systems -- qudits instead of qubits. {\it
  e.g.}
\cite{alber+01,stenbar98,albeverio+00,karimipour+02,grudwoj02}. In
this context attention was paid to the {\it d}--dimensional
generalization of Bell states. For any {\it d} it is possible to
construct $d^{2}$ mutually orthogonal maximally entangled states.
These states form a basis of the Hilbert space ${\cal H}_{d} \otimes
{\cal H}_{d}$, which will be called maximally entangled basis (MEB).
In a slight abuse of language, we use MEB to denote the plural form
maximally entangled bases, as well. The expansion of the acronym will
be clear from context.  The construction of MEB is, of course, not
unique and one can ask if there is any interesting classification of
the possible MEB of a given space. In this paper we propose to
classify the MEB with the use of the following notion of equivalence.
Two MEB are equivalent if and only if (iff) there exists a bilocal
unitary operation $U= U_{1} \otimes U_{2}$ which transforms all the
states of the first MEB into the states of the second one. Let us
denote the basis states of the two MEB as $|\Psi_{jk}^{(1)}\rangle $
and $|\Psi_{jk}^{(2)}\rangle $ $(j,k=0,1 \ldots, d-1)$.  Then these
two MEB are equivalent iff there exist permutations $\pi_{1}$ and
$\pi_{2}$, phases $\theta_{jk}$ and unitary operations $U_{1}$ and
$U_{2}$ such that
\begin{equation}
  |\Psi_{jk}^{(1)}\rangle = e^{i \theta_{jk}} (U_{1} \otimes U_{2})
  |\Psi_{\pi_{1}(j) \pi_{2}(k)}^{(2)}\rangle  
  \label{Rzoa}
\end{equation}
Of course it is always possible to locally transform a given maximally
entangled state into any other maximally entangled state. What we
require here is that all states of one basis be transformed into the
states of the second basis by the same unitary operation $U_{1}
\otimes U_{2}$. We will present an explicit construction of
inequivalent MEB of the Hilbert space ${\cal H}_{d} \otimes {\cal
  H}_{d}$, provided that {\it d} is not prime.

Let us begin by restating some basic properties of the well known
two-qubit Bell states
\begin{equation}
  |\Psi_{jk}^{{\rm Bell}}\rangle = \frac{1}{\sqrt{2}}
  \sum\limits_{m=0}^{1} (-1)^{jm} |m \rangle |m \oplus k \rangle ,
  \label{R2oa}
\end{equation}
where $\oplus$ denotes addition modulo $2$, while $j,k=0,1$. These
states can be generated from product states of two qubits by applying the
 \cnot gate ($\cnot |j \rangle |k \rangle = |j \rangle |k \oplus
j\rangle $) as in the following  equation
\begin{equation}
|\Psi_{jk}^{\rm Bell}\rangle = \cnot |\phi_{j}\rangle  |\chi_{k}\rangle ,
\label{R3oa}
\end{equation}
provided that the states $|\phi_{j}\rangle $ and $|\chi_{k}\rangle $
are appropriately chosen, namely $|\phi_{j}\rangle =
2^{-1/2}\sum_{m=0}^{1} (-1)^{jm}|m \rangle $ and
$|\chi_{k}\rangle = |k \rangle $.  The Hadamard gate $H$ ($H |j
\rangle = 2^{-1/2} \sum_{k=0}^{1} (-1)^{jk} |k \rangle) $ 
transforms computational basis states $|j \rangle $ into the states
$|\phi_{j}\rangle $, {\it i.e.} $|\phi_{j}\rangle = H |j \rangle $.
The Bell states clearly form an orthonormal basis $\{ |\Psi_{jk}^{\rm
  Bell}\rangle \}$, since they are generated from the orthonormal
basis $\{|j \rangle |k \rangle \}$ via the unitary transformation
$\cnot (H \otimes \openone)$.  In order to check if the states
obtained in this manner are really maximally entangled it suffices
(notice that whole system is in a pure state) to calculate the von Neuman
entropy of the reduced density operator of the first qubit
$S(\rho_{jk}^{A}) = - {\sf Tr}_{A} \rho_{jk}^{A} \log \rho_{jk}^{A}$,
where $\rho_{jk}^{A} = {\sf Tr}_{B} (|\Psi_{jk}\rangle_{AB}\langle
\Psi_{jk}|)$ for all $j,k$ (A and B denote the first and second qubit
respectively).  For maximally entangled states the von Neuman entropy
takes the value $S(\rho_{jk}^{A} =1)$.  It is easy to show that
$\rho_{jk}^{A} = 2^{-1} \sum_{m=0}^{1} |m \rangle \langle m|$ and
consequently $S(\rho_{jk}^{A}) =1$ indeed.  Thus, the Bell states form
a MEB in the Hilbert space ${\cal H}_{2} \otimes {\cal H}_{2}$.

In order to generalize the above procedure to the {\it d}--dimensional
case we will look for the set $\Phi =\{|\phi_{j}\rangle, j=0, 1,
\ldots, d-1\} $ of {\it d}--dimensional states $|\phi_{j}\rangle $
which can be used to construct MEB in the Hilbert space ${\cal H}_{d}
\otimes {\cal H}_{d}$ in the following way
\begin{equation}
|\Psi_{jk}^{\rm MEB}\rangle = \gnot |\phi_{j}\rangle  |k \rangle ,
 \; \; j,k= 0, 1, \ldots, d-1,
\label{R4oa}
\end{equation}
where the {\it d}--dimensional generalization of the \cnot gate is
defined as $\gnot |j \rangle |k \rangle = |j \rangle |j \ominus_{d}
k\rangle $, and  $\ominus_{d}$ denotes subtraction modulo $d$. We
follow the method suggested in \cite{alber+01} of generalizing the
\cnot gate with the use of subtraction instead of addition which leads
to the useful property of self--inverseness $\gnot^{2} =\openone$. The states
$|\Psi_{jk}^{\rm MEB}\rangle $ will form a MEB iff the states
$|\phi_{j}\rangle $ fulfil the following two conditions. First, the
states $|\phi_{j}\rangle $ have to form an orthonormal basis of ${\cal
  H}_{d}$.  Second, as was shown by Stenholm and Bardroff
\cite{stenbar98}, the condition
\begin{equation}
|\langle k| \phi_{j}\rangle | = \frac{1}{\sqrt{d}}
\label{R5oa}
\end{equation}
must hold for all $j,k$. This condition follows from the fact
that reduced density operator of the single qudit (for the two qudit
state $|\Psi_{jk}^{\rm MEB}\rangle $) is $\rho_{jk}^{A}= |\langle m|
\phi_{j}\rangle |^{2} \sum_{m=0}^{d-1} |m \rangle \langle m|$. Let us
denote by $V_{\phi } = \sum_{j} |\phi_{j}\rangle \langle j|$ a unitary
operation which transforms states $|j \rangle $ into $|\phi_{j}\rangle
$. From Eq.(\ref{R5oa}) one sees that all elements of the
corresponding matrix $(V_{\phi })_{jk} = \langle j|V_{\phi }|k \rangle
$ must have equal moduli $|(V_{\phi })_{jk}| = d^{- 1/2}$ .
Unitary matrices fulfulling these properties have been called
Zeilinger matrices \cite{stenbar98, torma+95}.  Two commonly used
operators belonging to this class are:
the {\it d}--dimensional discrete Fourier transform
\begin{equation}
  \dft_{d} = \sum\limits_{j,k} \omd^{jk} |j \rangle \langle k|,
  \label{R6oa}
\end{equation}
where $\omd = e^{i \frac{2 \pi }{d}}$ is the {\it d}--th complex root
of unity, and in the case when the dimension $d= 2^{n}$ is some
integer power of $2$, the generalized Hadamard operation $H_{n} $
defined recursively as follows
\begin{equation}
H_{n+1} = H_{1} \otimes H_{n}, \quad H_{1} = H.
\label{R7oa}
\end{equation}

In the qubit case these operations are identical, {\it i.e.} $\dft_{2}
= H$. This is not true, when $d=4$ where $\dft_{4} \neq H_{2}$ and
moreover, these operations will be shown later to generate inequivalent
MEB.  We say that the Zeilnger operation $V$ generates a MEB if the
states of this basis are given by
\begin{equation}
|\Psi_{jk}^{{\rm MEB}}\rangle = \gnot (V \otimes \openone) |j \rangle
 |k \rangle  
\label{R9oa}
\end{equation}

In order to present an explicit construction of MEB, we introduce a
function from the Cartesian product ${\cal H}_{d} \times {\cal H}_{d}$
to ${\cal H}_{d}$, which we will call vector multiplication.
Multiplication of vectors $|a \rangle = \sum_{j} a_{j} |j \rangle $
and $|b \rangle = \sum_{j} b_{j} |j \rangle $ gives a vector $|c
\rangle = |a \rangle \circ |b \rangle $ such that $|c \rangle =
\sum_{j} c_{j}|j \rangle $, where $c_{j} = \sqrt{d} a_{j} b_{j}$. It
is quite easy to prove that if the set $\Phi = \{ |\phi_{j}\rangle,
j=0,1, \ldots, d-1\} $ of {\it d} mutually orthogonal states
$|\phi_{j}\rangle $ together with the above defined multiplication
form a group $G=(\Phi ,\circ)$, then these states can be used to
construct  a MEB in accordance with Eq.(\ref{R4oa}). The order of the
group $G$ is $d$. Therefore any element $|\phi_{j}\rangle $ of group
$G$ must fulfil
\begin{equation}
(|\phi_{j}\rangle )^{\circ d} = |1_{G}\rangle,
\label{R0oa}
\end{equation}
where $|1_{G}\rangle $ denotes the unit element of the group. Clearly,
$|1_{G}\rangle = d^{- 1/2}\sum_{j} |j \rangle $. On the other
hand, by the definition of the group operation $ (|\phi_{j}\rangle
)^{\circ d} = \sum_{k} d^{(d-1)/2} \langle k|
\phi_{j}\rangle^{d} |k \rangle $. Thus Eq.(\ref{R0oa}) leads to the
identity $(d^{1/2} \langle k| \phi_{j}\rangle )^{d} =1$ and
consequently to $\langle k| \phi_{j}\rangle  = d^{-
  1/2}\omd^{m}$ for some integer $m$. One can thus see that
the condition of Eq.(\ref{R5oa}) is automatically fulfilled for the
elements of the group $G$.  Let us now present an explicit
construction of the elements $|\phi_{j}\rangle $ of group $G$. $G$ is
a finite Abelian group, so it is either a cyclic group $G_{d} $ or a
direct product of cyclic groups $G_{d_{1}} \times G_{d_{2}} \times
\ldots G_{d_{r}}$($G_{k}$ denotes the cyclic group of order $k$). In
the former case we can define inequivalent characters $\chi^{(n)}$
$(n=0,1, \ldots d-1)$ of irreducible representations of the group
$G_{d}$ as $\chi^{(n)} (|\phi_{j}\rangle ) = \omd^{nj}$. The
characters fulfil the orthogonality relation \cite{cantrell},
$\sum_{j=0}^{d-1} \chi^{(n)}(|\phi_{j}\rangle) \chi^{(m)} (|\phi_{j}\rangle
)^{\ast} = d \delta_{nm}$.  Thus, if we take  vectors
$|\phi_{j}\rangle $ of the form 
\begin{equation}
|\phi_{j}\rangle  = d^{- \frac{1}{2}} \sum\limits_{k=0}^{d-1}
 \chi^{(j)} (|\phi_{k}\rangle ) |k \rangle ,
\label{Rapa}
\end{equation}
they will form an orthonormal basis $\langle \phi_{j} |
\phi_{k}\rangle = \delta_{jk}$. It should be emphasized here that the
structure of the group does not change under the permutation of the
computational basis states. Thus, there exist $d!$ bases given by
$|\phi_{j}^{\pi }\rangle = d^{- 1/2} \sum_{j=0}^{d-1}
\chi^{(j)} (|\phi_{k}\rangle ) |\pi (k) \rangle $, where $\pi $ is an
arbitrary permutation.  On the other hand, if $G$ is a direct product
of cyclic groups, Eq.(\ref{Rapa}) can be also used to obtain an
alternative construction of vectors $|\phi_{j}\rangle $ provided that
we use the characters of the irreducible representations of group $G$
nonisomorphic with $G_{d}$. This is possible if the order $d$ of the
group can be expressed as a nontrivial product of integers $d=d_{1}
d_{2} \ldots d_{r}$. Any such decomposition of $d$ leads to group $G$
of the form $G= G_{d_{1}} \times G_{d_{2}} \times \ldots \times
G_{d_{r}}$. For each $G_{d_{i}}$ we define $d_{i}$ inequivalent
characters of its irreducible representation
$\chi_{i}^{n_{i}}(|\phi_{j}\rangle )= \omega_{d_{i}}^{n_{i} m_{i}}$
$(n_{i}, m_{i} = 0,1, \ldots, d_{i}-1)$, where the one--to--one
mapping between indices $j$ $(j=0,1, \ldots, d-1)$ and $\{ m_{i}, i=1,
\ldots, r \}$ given by $j= \sum_{i=1}^{r} m_{i} \delta_{i}$, where
$\delta_{i}= d (\Pi_{k=i}^{r} d_{k})^{-1}$ is used. We can now obtain
$d$ characters of the irreducible representations of group $G$ as a
product of characters of groups $G_{d_{i}} $, {\it i.e.}
$\chi^{(n)}(|\phi_{j}\rangle) = \Pi_{i=1}^{r}
\chi^{(n_{i})}_{i}(|\phi_{j}\rangle )$, where $n= \sum_{i=1}^{r} n_{i}
D _{i}$ and $D_{i} = \Pi_{k=i+1}^{r} d_{k}$. In this way we obtain
\begin{equation}
  \chi^{(n)} (|\phi_{j}\rangle ) = \omd^{\vec{n} \cdot \vec{j}},
  \label{Rbpa}
\end{equation}
where we introduce the product $\vec{n} \cdot \vec{j} = d
\sum_{i=1}^{r} \frac{n_{i}m_{i}}{d_{i}}$ dependent on the
decomposition $d=d_{1} d_{2} \ldots d_{r}$.  Eqs.(\ref{Rapa}) and
(\ref{Rbpa}) lead to construction of nonisomorphic groups of vectors
$|\phi_{j}\rangle $ for each decomposition of $d$ with the use of
operator $V$ $(V |j \rangle = |\phi_{j}\rangle )$ given by
\begin{equation}
  V = d^{- \frac{1}{2}} \sum\limits_{j,k=0}^{d-1} \omd^{\vec{k} \cdot
    \vec{j}} |k \rangle \langle j|.
  \label{Rcpa}
\end{equation}

The operators $\dft_{4}$ and $H_{2}$, mentioned earlier, are two
examples of operators constructed in this way for the two possible
decompositions of $d=4$, namely $4=4$ and $4=2 \times 2$. 

Let us now consider two operators and given by Eq.(\ref{Rcpa}) with
the use of different decomposition of $d$ and two MEB generated by these
operators
\begin{subequations}
\label{Rdpa}
\begin{eqnarray}
|\Psi_{jk}^{(1)}\rangle  = \gnot (V_{1} \otimes \openone) |j \rangle
||k \rangle , \\
|\Psi_{jk}^{(2)}\rangle  = \gnot (V_{2} \otimes \openone) |j \rangle
||k \rangle  .
\end{eqnarray}
\end{subequations}
These equations lead to the following transformation 
\begin{equation}
|\Psi_{jk}^{(1)}\rangle  = \gnot(V_{1} P_{1}^{-1} V_{2}^{-1} \otimes
 P_{2}^{-1}) \gnot |\Psi_{\pi (j) \pi (k)}^{(2)}\rangle 
\label{Rgpa}
\end{equation}
where $P_{1}$ and $P_{2}$ are permutation operators corresponding to
permutations $\pi_{1}$ and $\pi_{2}$ . Thus two MEB given by
Eq.(\ref{Rdpa}) are equivalent (see Eq.(\ref{Rzoa})) iff there exist
local operators $U_{1}$ and $U_{2}$ such that
\begin{equation}
U_{1} \otimes U_{2} = \gnot (V_{1} P_{1}^{-1} V_{2}^{-1} \otimes
P_{2}^{-1}) \gnot.
\label{Rhpa}
\end{equation}
This is possible iff the operator $V_{1} P_{1}^{-1} V_{2}^{-1}$ is the
product of the permutation operator $P$ and unitary diagonal operator
$D$, {\it i.e.} $V_{1} P_{1}^{-1} V_{2}^{-1}= PD$ , which leads to the
following condition of equivalence of two MEB $|\Psi_{jk}^{(1)}\rangle
$ and $|\Psi_{jk}^{(2)}\rangle $
\begin{equation}
P^{-1} V_{1} P_{1}^{-1} = DV_{2}.
\label{Ripa}
\end{equation}
Let us notice that due to group structure of the columns of matrix
$V_{1}$, one of these columns must be of the form 
\begin{equation}
\left(
\begin{array}{c}
     d^{- \frac{1}{2}}\\
     d^{- \frac{1}{2}}\\
     \vdots \\
     d^{- \frac{1}{2}}
\end{array}
\right)
\label{Rjpa}
\end{equation}
Obviously one of the columns of matrix $V_{1} P_{1}^{-1} V_{2}^{-1}$
must also have this form. This restricts the possible form of diagonal
matrix $D$, namely $D_{jj}= (V_{2})_{jk}^{\ast}$ must hold for some
$k$.  On the other hand, the group structure guarantees that there
exist $k'$ such that $(V_{2})_{jk}^{\ast} = (V_{2})_{jk'}$. It means
that the action of $D$ on $V_{2}$ is equivalent to performing vector
multiplication $\circ$ of the columns of $V_{2}$ by one of these 
columns, {\it i.e.}  equivalent to permutation of the columns of
$V_{2}$. This leads to the final conclusion that two operators $V_{1}$ and
$V_{2}$ given by Eq.(\ref{Rcpa}) generate MEB, which are equivalent
iff they are equal to each other up to permutations of columns and
rows. It follows that two operators $V_{1}$ and $V_{2}$ given by
Eq.(\ref{Rcpa}) with the use of different decomposition of $d$
generate inequivalent MEB.  As an example we can take MEB generated by
$\dft_{4}$ and $H_{2}$ and conclude that these MEB are inequivalent.

In conclusion, we have introduced the notion of equivalence of
maximally entangled bases of qudits. In accordance with this notion,
we have presented an explicit construction of inequivalent maximally
entangled bases from group--theoretic concepts.

We would like to thank the State Committee for Scientific Research for
financial support under grant no. 0 T00A 003 23.


\end{document}